\documentclass[12pt, oneside]{article}   	% use "amsart" instead of "article" for AMSLaTeX format
\usepackage{geometry}                		% See geometry.pdf to learn the layout options. There are lots.
\usepackage{graphicx}
\usepackage{epstopdf}
\usepackage{graphicx}				% Use pdf, png, jpg, or eps§ with pdflatex; use eps in DVI mode
\usepackage{amsmath}								% TeX will automatically convert eps --> pdf in pdflatex		
\usepackage{amssymb}
\usepackage{color}
\usepackage[usenames,dvipsnames,svgnames,table]{xcolor}
\usepackage[colorlinks=true,
            linkcolor=red,
            urlcolor=magenta,
            citecolor=blue]{hyperref}
\title{Energy spreading, equipartition and chaos in lattices with non-central forces }

\author{\fontsize{10}{10}\selectfont Arnold Ngapasare$^{1}$, \ Georgios Theocharis$^{2}$, \ Olivier Richoux$^{2}$\\
\fontsize{10}{10}\selectfont \ Vassos Achilleos$^{2}$ \ and \ Charalampos Skokos$^{1}$ \\
\fontsize{10}{10}\selectfont $^{1}${Nonlinear Dynamics and Chaos Group} \\
\fontsize{10}{10}\selectfont {Department of Mathematics and Applied Mathematics} \\  
\fontsize{10}{10}\selectfont {University of Cape Town, Rondebosch 7701, South Africa. } \\  % The line break was forced via \\
\fontsize{10}{10}\selectfont $^{2}${Laboratoire d'Acoustique de l'Universit\'{e} du Mans}\\
\fontsize{10}{10}\selectfont {UMR CNRS 6613 Av. O.~Messiaen, F-72085 LE MANS Cedex 9, France.}}   % The line break was forced via \\
\date{}							% Activate to display a given date or no date

\begin{document}
\maketitle

\begin{abstract}
{\fontsize{10}{10}\selectfont 
We numerically study a one dimensional, nonlinear lattice model which in the linear limit is relevant to the study of bending (flexural) waves. In contrast with the classic one dimensional mass-spring system, the linear dispersion relation of the considered model has different characteristics in the low frequency limit. By introducing disorder in the masses of the lattice particles, we investigate how different nonlinearities (cubic, quartic and their combination) lead to energy delocalization, equipartition and chaotic dynamics. We excite the lattice using single site initial momentum excitations corresponding to a strongly localized linear mode and increase the initial energy of excitation. Beyond a certain energy threshold, when the cubic nonlinearity is present, the system is found to reach energy equipartition and total delocalization. On the other hand, when only the quartic nonlinearity is activated, the system remains localized and away from equipartition at least for the energies and evolution times considered here. However, for large enough energies for all types of nonlinearities we observe chaos. This chaotic behavior is combined with energy delocalization when cubic nonlinearities are present, while the appearance of only quartic nonlinearity leads to energy localization. Our results reveal a rich dynamical behavior and show differences with the relevant Fermi-Pasta-Ulam-Tsingou model. Our findings pave the way for the study of models relevant to bending (flexural) waves in the presence of nonlinearity and disorder, anticipating different energy transport behaviors.}
\end{abstract}

%\textbf{Keywords:} Anderson Localization, Energy spreading, Energy equipartition, Chaos

\section{Introduction} \label{sec1}
The work of P.W.~Anderson on wave dynamics in disordered systems has been hailed as one of the major milestones of $20$th-century physics \cite{1}. Since the publication of that pioneering work, several studies of Anderson localization (AL), which is the exponential localization of almost all eigenmodes of random realizations of mass and/or force constant distributions, have been performed \cite{7,4,6,8,3,9,10,5,2}. The localization of wave functions (or eigenmodes) leads to the absence of energy transport in disordered systems. Over the years, AL has been extended to other models, apart from the original electronic systems, and related phenomena continue to reveal interesting results across many branches of physics \cite{11,12,13,14,15,16}. Another scientific landmark of the previous century, which has culminated in a whole new branch of physics, is the work of Enrico Fermi and his collaborators \cite{17,18}. Indeed, their work on the now famous Fermi-Pasta-Ulam-Tsingou (FPUT) problem gave birth to what we now know as computational physics [see Ref.~\cite{19} and references therein]. 

There are two main mechanisms through which waves can be localized in lattice chains, namely (i) disorder and (ii) nonlinearity, resulting in interesting physical phenomena such as AL \cite{1}, solitons \cite{20} and discrete breathers \cite{21,22}. Solitons are shape preserving propagating localized formations, while discrete breathers are spatially localized time periodic structures. The interplay of nonlinearity and disorder has been an area of intense active research over the years (see Ref.~\cite{23} and references therein for a recent review on the topic of disorder and nonlinearity). In earlier studies, the actual role of nonlinearity on disorder remained not so clear. However, nowadays the theory of the interplay of disorder and weak nonlinearity in lattices such as the disordered  Klein-Gordon (DKG) lattice of anharmonic oscillators and the disordered discrete nonlinear Schr\"{o}dinger equation (DDNLS) has been well developed \cite{24,60}. It is now known that weak nonlinearity leads to subdiffusive energy transport in these models. On the other hand, recent studies of discrete nonlinear models describing mechanical lattices have exploited the strongly nonlinear limit and have revealed superdiffusive energy transport in the presence of disorder \cite{15,25,26}.

Regarding mechanical lattices (see e.g., Refs.~\cite{33,62}), one of the most fundamental models describing the longitudinal displacement of point masses connected with springs is the FPUT model \cite{17,18}. For the linear regime of this model it is known that energy propagation in the presence of disorder is dictated by the form of the dispersion relation in the low frequency limit which starts from zero with a linear slope \cite{63}. The nonlinearity is then found to generally, enhance energy transport although a clear understanding of the mechanisms is not yet established \cite{43,57}. Here, we wish to extend these studies, by considering a discrete model of a multidimensional conservative Hamiltonian system featuring a different dispersion relation in the low frequency regime. The proposed model is inspired by the recent interest on incorporating other degrees of freedom in mechanical lattices (e.g., rotations \cite{34}) and also on the idea of AL of flexural waves \cite{61}. We study a toy model whose linear limit describes lattice bending vibrations. Our main goal is to explore the interplay of nonlinearity and disorder and identify differences in energy transport with other models (e.g., the DKG and/or the FPUT models), which feature different linear spectra \cite{23,24,25}.
In our investigations we consider single site excitations on lattice sites which are not too close to the boundaries. This excitation choice typically results in the excitation of a single localized mode, due to the presence of strong mass-disorder. In particular, we endeavor to answer the question of the conditions under which the energy initially given to a single mode at a single initial excitation will spread to the rest of the lattice and whether the system will eventually reach energy equipartition or not, as well as study its chaotic dynamics. 
Here we focus on the chaotic behavior of conservative lattice systems, but the role of chaos in various setups of continuous and discrete time models is also a field of active research (see e.g., Refs.~\cite{71,72,73,74}).

The rest of this paper is organized in the following manner. In Section~\ref{sec2}, we describe the proposed one-dimensional ($1$D) discrete model. In the next two sections, we discuss some representative results about the effects of a momentum single site initial excitation on energy localization, equipartition (Section~\ref{sec3}) and chaos (Section~\ref{sec4}). In Section~\ref{sec5}, we present a statistical analysis of the localization, equipartition and chaotic behaviors of the model for a range of energy values. Finally, in Section~\ref{sec6}, we summarize our work and present its conclusions.

\section{The Hamiltonian model and its equations of motion} \label{sec2}

\begin{figure}[ht!]
\begin{center}
 \includegraphics[width=0.7\textwidth,height=0.7\textheight,keepaspectratio]{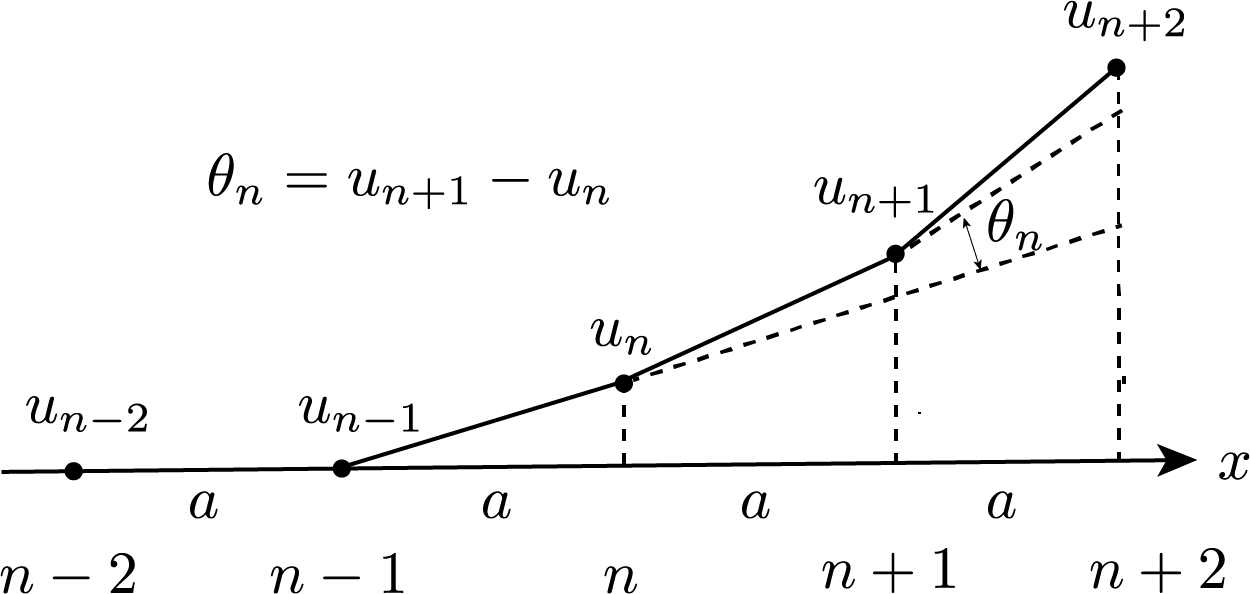}   
\caption{A schematic representation showing the relation between the rotation angles $\theta_{n}$, and the relative transverse displacement of the point mass at $u_n$ of a lattice with distance $a$ between neighboring masses (in our study we set $a=1$). In the small angle approximation the equations of motion for the lattice is the linear limit of the Hamiltonian (\ref{eqH1}).}
\label{fig1}
\end{center}
\end{figure}
We consider a discrete nonlinear model  described by the following  Hamiltonian 
\begin{equation} \label{eqH1}
\begin{aligned}
H_{34}=\sum_{n=1}^{N}H^{(n)}= \sum_{n=1}^{N} \frac{P_n^2}{2m_n}+\frac{K_2~\zeta^2_n}{2}+\frac{K_3 ~\zeta^3_n}{3}+\frac{K_4~\zeta^4_n}{4}, 
\end{aligned}
\end{equation}
where $\zeta_n =(u_{n+1}-u_n)- (u_{n}-u_{n-1})$  and $P_n=m_n\dot{u}_n$. %denotes respectively, the model's relative displacements at neighbouring site pairs and momentum of the $n$th particle.
In our notation $[~\dot{}~]$ denotes derivative with respect to time and $m_n$ corresponds to the mass of the $n$th site. In the linear limit, i.e., when  $K_{3}=K_{4}=0$,  Eq.~(\ref{eqH1}) describes the transverse displacements of a mass spring chain as illustrated in Fig.~\ref{fig1}~\cite{41}. It is worth noting from the sketch in Fig.~\ref{fig1} that for sufficiently small rotation angles $\theta_n$, the relation $\theta_n-\theta_{n-1}\equiv\zeta_n= \dfrac{1}{a} (u_{n+1}-u_n)- \dfrac{1}{a}(u_{n}-u_{n-1})$ holds, with $a$ being the distance between neighboring sites. For simplicity in our study we set $a=1$. 

For the purposes of our work, we consider three different cases regarding the nonlinear terms. One case has only cubic nonlinearity $(K_4=0)$ which we refer to as $H_3$, an additional case with only quartic nonlinearity $(K_3=0)$ is, from now on, referred to as $H_4$, and finally the general case where both nonlinear terms are present ($K_3 \neq 0$, $K_4 \neq 0$), which we refer to  as $H_{34}$.  Note also that in this work $K_2>0$ and we consider disorder only in the masses. The coefficients $K_{i}$, $i=3,4$ (when non-zero) are kept constant and the corresponding equations of motion for the general $H_{34}$ model are
\begin{equation} \label{eq23}
\begin{aligned}
 m_n  \ddot{u}_{n} &  =   2 K_2 \Big [ (u_{n+1}-u_{n})-(u_{n}- u_{n-1})\Big ]
 - K_2 \Big [ (u_{n}-u_{n-1})-(u_{n-1}- u_{n-2})\Big ]\\ 
 &- K_2 \Big [ (u_{n+2}-u_{n+1})-(u_{n+1}- u_{n})\Big ] 
  + 2 K_3 \Big [ (u_{n+1}-u_{n})-(u_{n}- u_{n-1})\Big ]^2 \\
 & - K_3 \Big [ (u_{n}-u_{n-1})-(u_{n-1}- u_{n-2})\Big ]^2 
 - K_3 \Big [ (u_{n+2}-u_{n+1})-(u_{n+1}- u_{n})\Big ]^2  \\
 & + 2 K_4 \Big [ (u_{n+1}-u_{n})-(u_{n}- u_{n-1})\Big ]^3 
  - K_4 \Big [ (u_{n}-u_{n-1})-(u_{n-1}- u_{n-2})\Big ]^3 \\
  & - K_4 \Big [ (u_{n+2}-u_{n+1})-(u_{n+1}- u_{n})\Big ]^3. 
\end{aligned}
\end{equation}

\subsection{Dispersion relation and eigenmodes}
\begin{figure}[b!]
\begin{center}
  \includegraphics[width=0.45\textwidth,height=0.45\textheight,keepaspectratio]{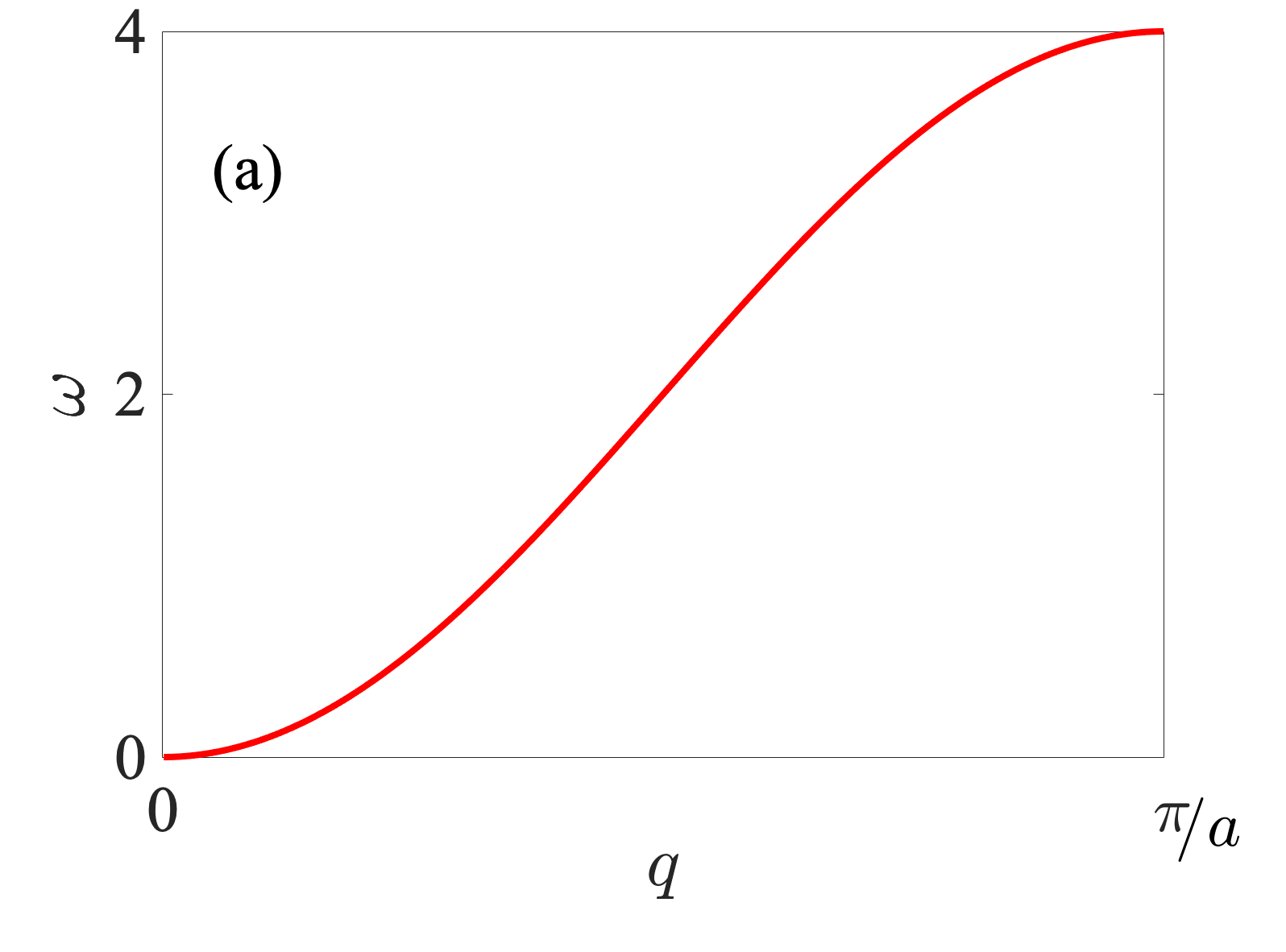} 
  \includegraphics[width=0.45\textwidth,height=0.45\textheight,keepaspectratio]{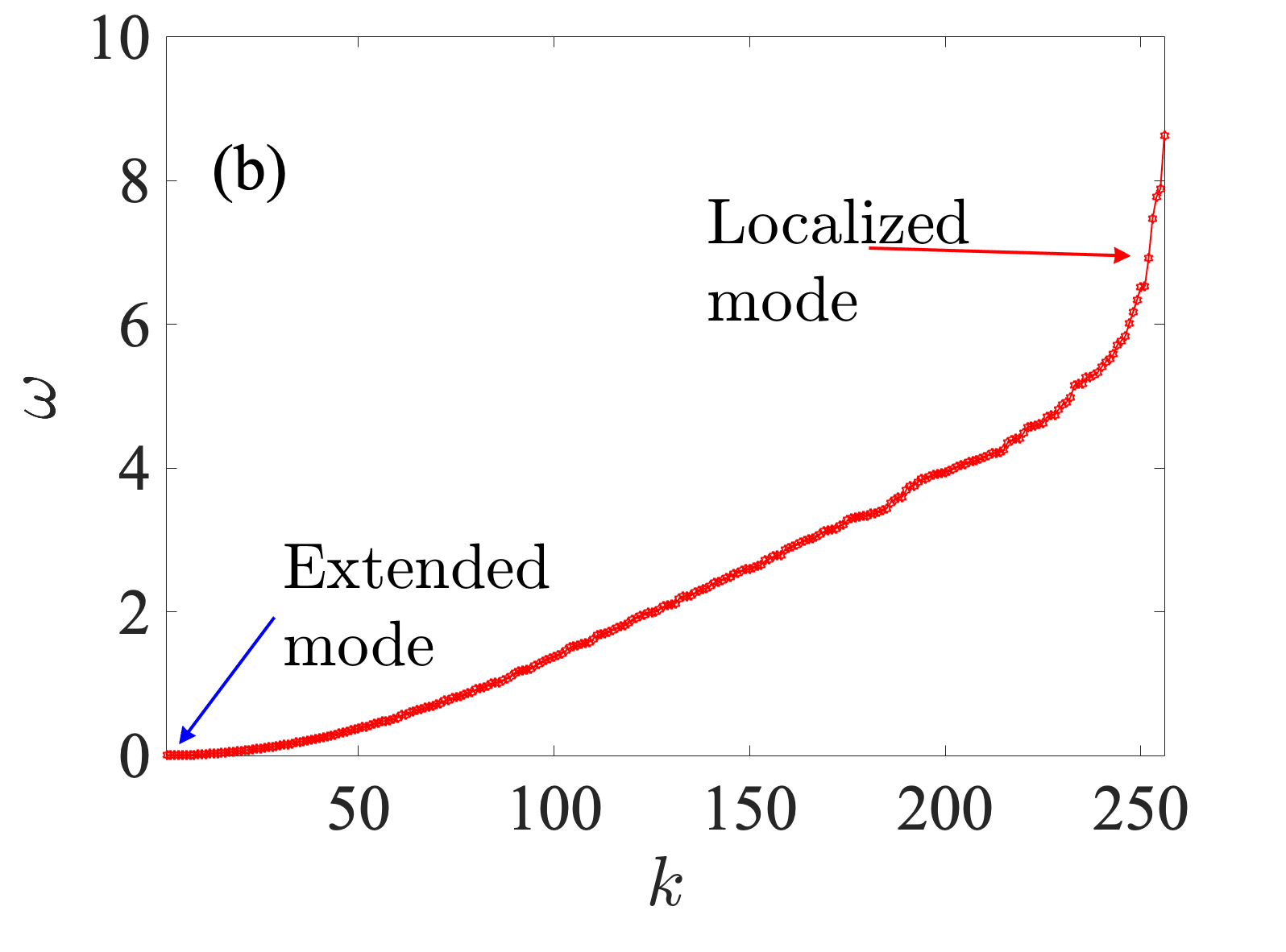}  
\end{center}
\caption{(color online) (a) Dispersion relation~(\ref{eq45}) of the lattice bending waves where $a$ is the lattice spatial constant (which is set to $a=1$ in our study), $q$ the wave number and $\omega$ the frequency. (b) The eigenfrequency spectrum of a representative disorder chain of size $N= 2^8$ with the arrows indicating eigenfrequencies for an extended (blue arrow) and a localized (red arrow) normal mode. The eigenfrequencies have been sorted and indexed from lowest to highest frequency.}
\label{fig2}
\end{figure}
Firstly, let us consider some aspects of the linear periodic case   corresponding to $K_3 = K_4 = 0$ and $m_n=m$.  We assume Bloch-like solutions of Eq.~(\ref{eq23}) of the form $ u_n = A e^{i(qn -\omega t)} $. This solution models traveling waves in a lattice chain, so that each particle oscillates about its equilibrium position with the same frequency $\omega$, and amplitude $A$, while $q$ denotes the Bloch wave number. Using this ansatz we obtain the system's dispersion relation, which is given by the following equation
\begin{equation} \label{eq45}
\omega = \sqrt{ \frac{4 K_2 }{m} \Big [ 4 \sin ^2 (q/2)  - \sin ^2 q   \Big ] },
\end{equation}
and is also plotted in Fig.~\ref{fig2}(a). The dispersion relation of this  model is quite different from that of the harmonic chain with central forces, leading to some interesting dynamics. The uniqueness stems from the particularity of zero group velocity in the low frequency regime. Note that for low frequencies the dispersion relation of Eq.~(\ref{eq45}) resembles the one of a wave equation with  fourth order spatial derivative describing bending waves in elastic beams \cite{41}.

In what follows we consider finite chains with a size of  $N=2^8$ particles. In addition, we introduce disorder in the particle masses such that the random masses are taken from a uniform distribution, i.e., $m_n \in [ 0.1,1.8]$ while we fix the linear potential coefficient to $K_2=1$. The nonlinear coefficients $K_{3}$, $K_{4}$, take values of either $1$ or $0$ such that whenever any of the nonlinear coefficients is not zero, it is then set to $1$, e.g., for the system $H_4$ we set  $K_3=0$ and $K_4 = 1$. In \cite{43} it was shown that the results do not significantly differ if either mass or stiffness disorder is implemented hence for this reason we will consider only the case of mass disorder. 

We now describe the localization properties of the linear normal modes of this system. For this purpose, we consider solutions of the linear part of the system of $N$ masses given in Eq.~(\ref{eq23}) to be of the form $\mathbf{u}(t) = \mathbf{U}e^{-i\omega t}$, where $\mathbf{U} = [U_1, U_2, \dots ,U_N]^T$ with $(^{T} )$ denoting the transpose of a matrix. Hence, we obtain the eigenvalue problem
\begin{eqnarray}
-\omega^2_k \mathbf{M} \mathbf{U}_k = \mathbf{K} \mathbf{U}_k, \; k=1,...,N,
\end{eqnarray}
where $\omega_k$ are the eigenfrequencies and $\mathbf{U}_k$ the corresponding eigenvectors. $\mathbf{M}$ is a diagonal matrix containing the mass elements, whilst $ \mathbf{K}$ is a sparse diagonal matrix of the coefficients. We show an eigenfrequency spectrum for a representative disordered chain of size $N = 2^8$ in Fig.~\ref{fig2}(b). First we observe that the very low frequency part of the spectrum is similar to the ordered case (where all masses are equal $m_n=m=1$), while the upper part has changed significantly and the cutoff frequency has been increased.

\begin{figure}[h!]
\begin{center}
\includegraphics[width=0.45\textwidth,height=0.45\textheight,keepaspectratio]{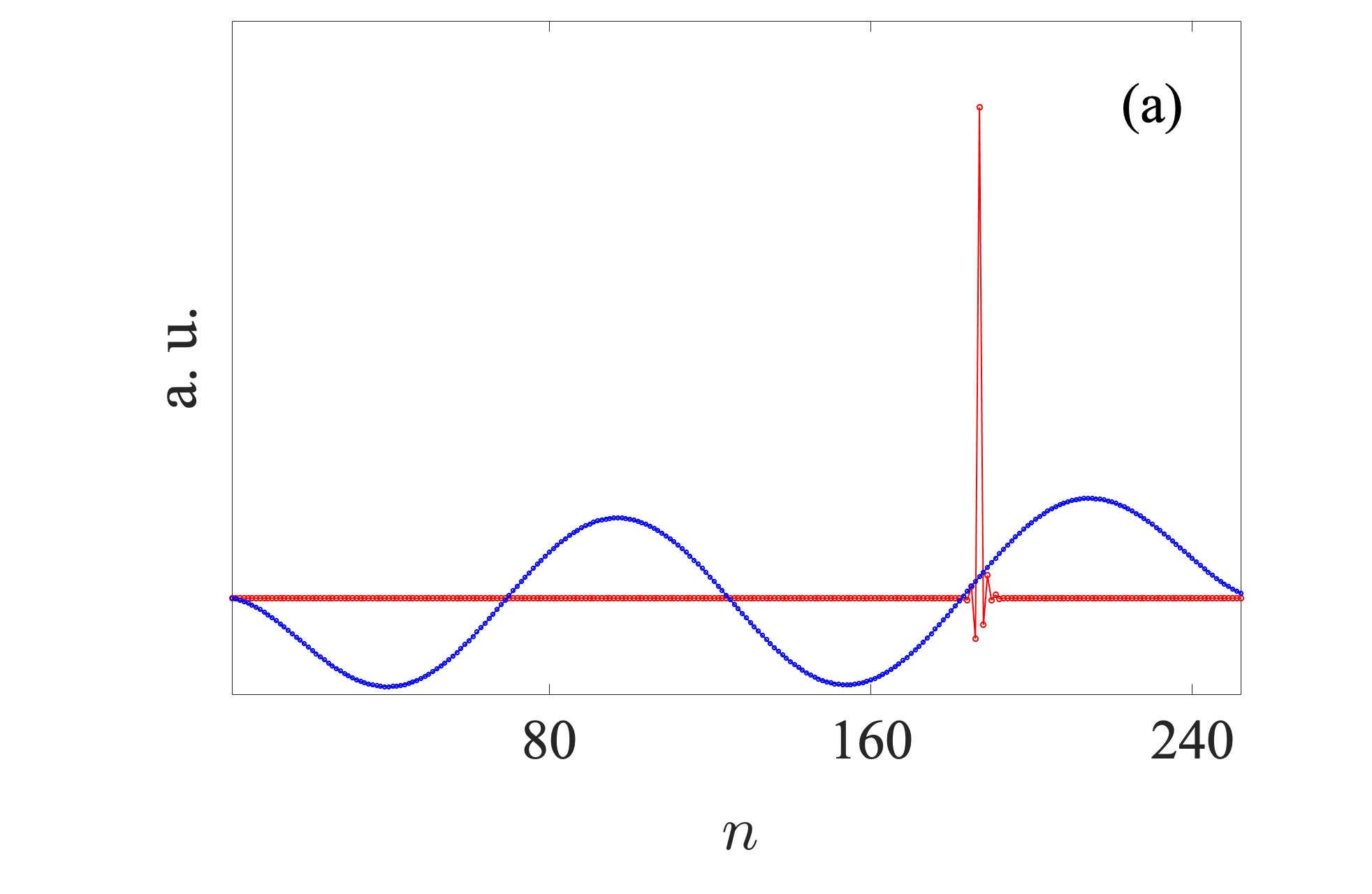} 
\includegraphics[width=0.45\textwidth,height=0.45\textheight,keepaspectratio]{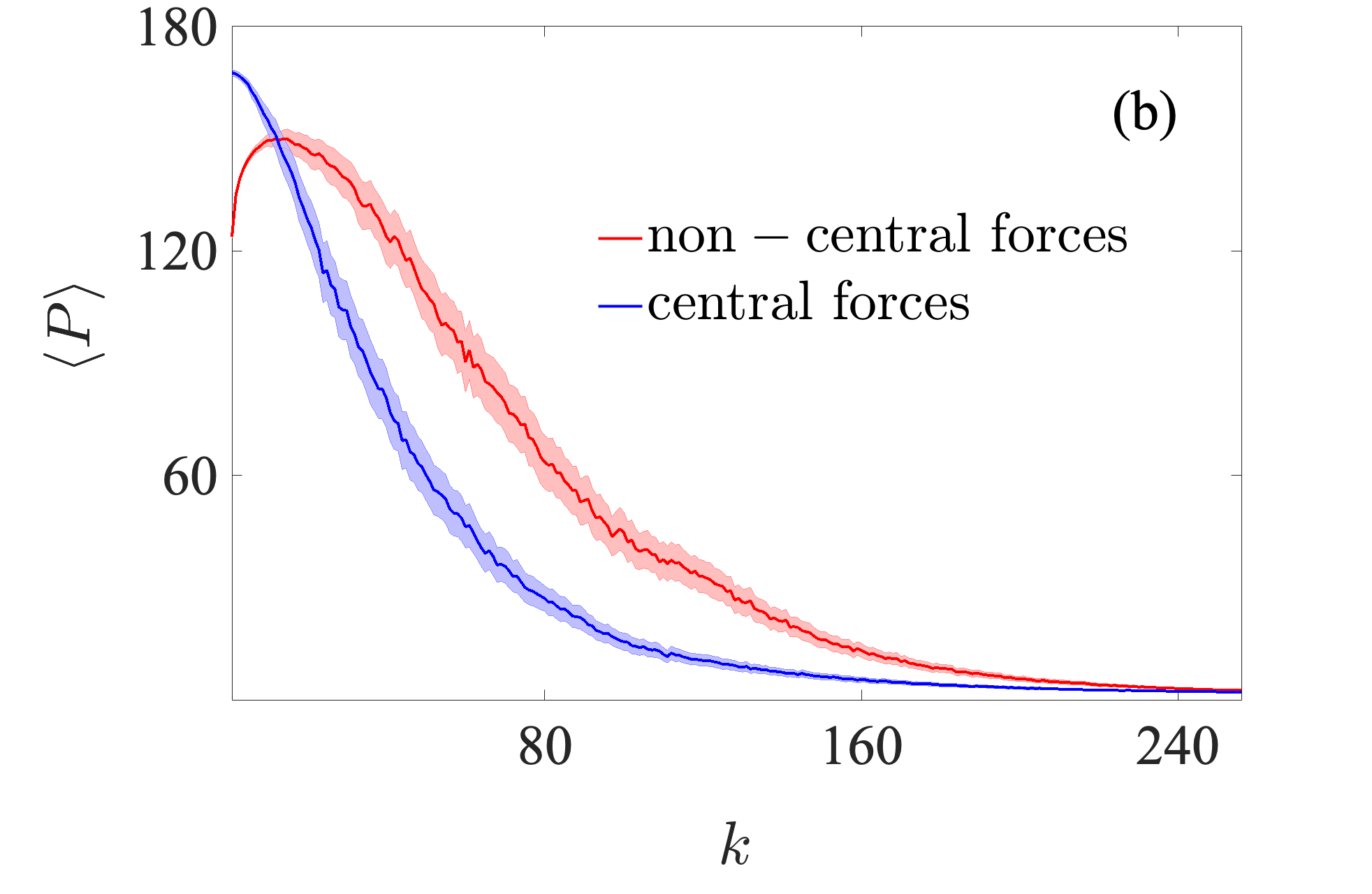} 
\end{center}
\caption{(color online) (a) Profiles of an extended (blue curve), as well as a localized (red curve) normal mode respectively indicated by the blue and red arrows in Fig.~\ref{fig2}(b), for a mass disordered harmonic chain with non-central forces. (b) Mean participation number $\langle P \rangle$ over $500$ disorder realizations of linear eigenmodes for harmonic chains with non-central forces (red curve) and harmonic chains with central forces (blue curve) of size $N = 2^8$. In (b), the modes have been sorted and indexed from the lowest frequency to the highest frequency modes. One standard deviation in the computation of $\langle P \rangle$ is indicated as the lightly shaded areas in (b).}
\label{fig3}
\end{figure}

Next, we are interested to characterize the localization properties of the modes which can be done by computing the participation number

\begin{equation} \label{eqP}
P = 1/ \sum_{n=1}^N h_n^2 ,
\end{equation}
where $h_n = H_n / H$ is the normalized  energy of each particle (see e.g., \cite{24} for a further discussion of this quantity). First we note that,
as is the case for the classic mass-spring model with central forces with disorder \cite{28}, the lower part of the spectrum comprises of extended modes  while the upper part of mostly localized ones. To illustrate that, in Fig.~\ref{fig3}(a) we plot the profiles  for the two characteristic modes indicated in Fig.~\ref{fig2}(b).
To have a more global picture and for the sake of  comparison, in Fig.~\ref{fig3}(b) we also plot the averaged over $500$ disorder realizations, mean participation number $\langle P \rangle$, for linear chains of two types. In particular we contrast for the same disorder parameters (the random masses are taken from a uniform distribution, i.e., $m_n \in [ 0.1,1.8]$) the harmonic chains with central forces, for which the equations of motion become $m_n \ddot{u}_n = u_{n+1} + u_{n-1} - 2 u_{n} $ (blue curve) and non-central forces obtained by setting $K_3=K_4=0$ in Eqs.~(\ref{eqH1}) and (\ref{eq23}) (red curve). We note some of the most glaring differences which are: (a) modes with the lower indices ($k \lesssim 20$) have higher $\langle P \rangle $ for the chain with central forces than for the chain with  non-central forces and (b) the chain with central forces has more localized modes as compared to the chain with  non-central forces at any lattice size [$\langle P \rangle \approx 2 $ for $n \gtrsim 160 $ in Fig.~\ref{fig3}(b)].

\section{Evolution of linear modes, energy delocalization and equipartition} \label{sec3}

\begin{figure}[ht!]
\begin{center}
\includegraphics[width=0.65\textwidth,height=0.65\textheight,keepaspectratio]{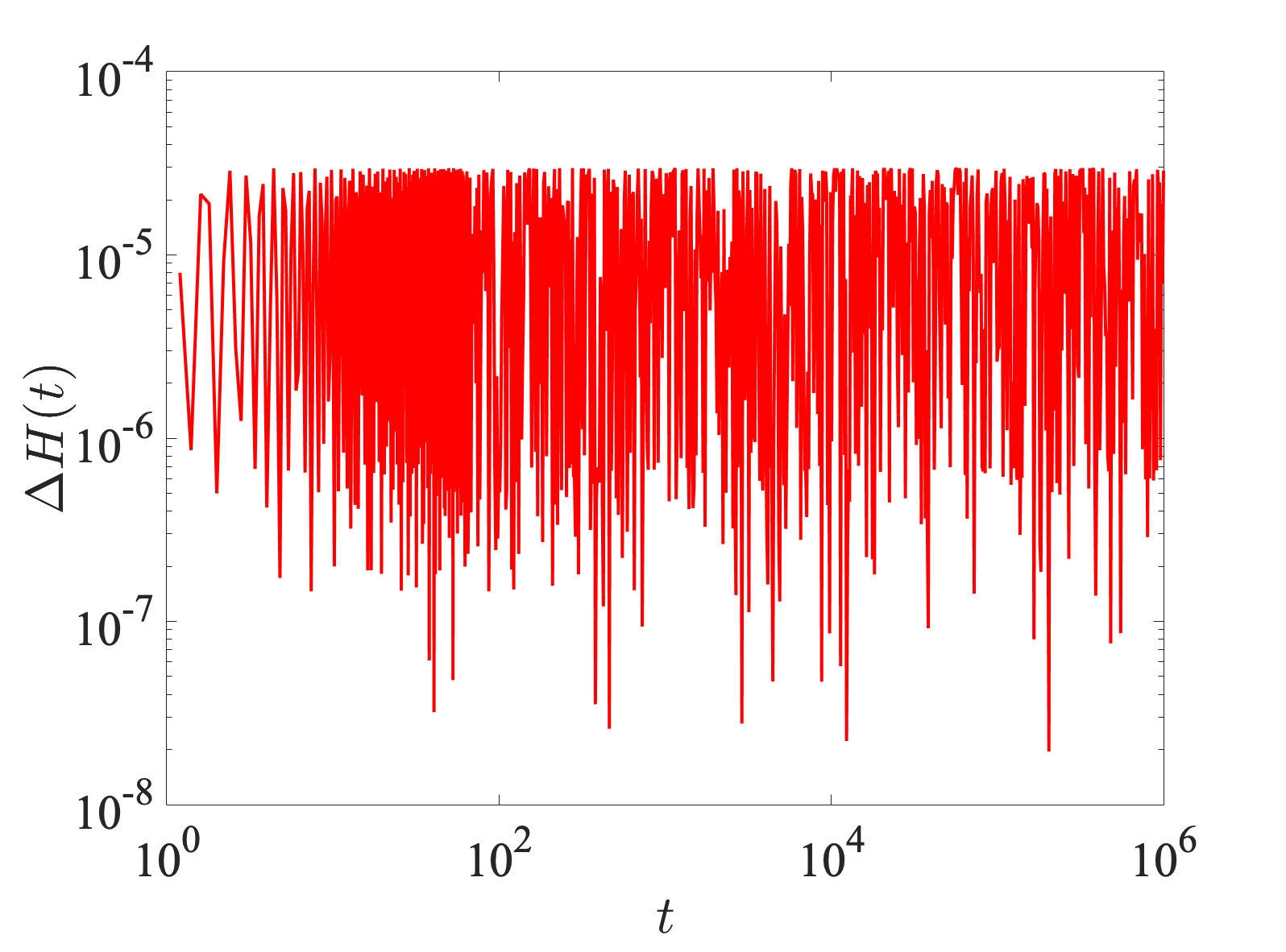}   
\end{center}
\caption{(color online)  Temporal evolution of the relative energy error $\Delta H (t) $ \eqref{eq:relative_error} of a single disordered realization with energy $H_4 = 0.2$ and time step $ \tau = 0.1$. }
\label{fig4}
\end{figure}

In this section, we consider the fate of a single site momentum initial excitation under the influence of the different kinds of nonlinearities (see the three different kinds of Hamiltonians presented in Section~\ref{sec2}). In Fig.~\ref{fig3}(a) we depict, using a red curve, an almost single site localized mode, namely mode $k=252$, for which only site $n=187$ is significantly displaced from its equilibrium position. Note that the size $N$ of the lattice determines the number of highly localized modes. In our study, we set the size of the chain to be $N =2^8$ in order to have a significant number of highly localized modes and implement fixed boundary conditions ($u_0 = u_{n+1} =  \dot{u}_0 = \dot{u}_{n+1} = 0$). The dynamics is studied by numerically solving the different types of equations of motion given in Section~\ref{sec2} using the \texttt{ABA864} symplectic integrator \cite{44,45,46}. This integration scheme allows the accurate conservation of the total energy $H$, and keeps the relative energy error \begin{equation}
\Delta H (t)  = \left|\dfrac { H (t) - H (0)  }{ H (0) }\right| 
\label{eq:relative_error}
\end{equation} 
less than $10^{-4}$ when the integration time step is $\tau = 0.1$. For illustrative purposes, we show in Fig.~\ref{fig4} a representative example of the  time evolution of $\Delta H (t)$ for a case of the $H_4$ system. From the presented results it is clearly seen that the used integration scheme keeps the error bounded for the whole duration of the simulation and that its upper bound does not increase in time.

\subsection{Spatiotemporal evolution of modes}
\begin{figure}[b!]
\begin{center}
\includegraphics[width=0.9\textwidth,height=0.5\textheight,keepaspectratio]{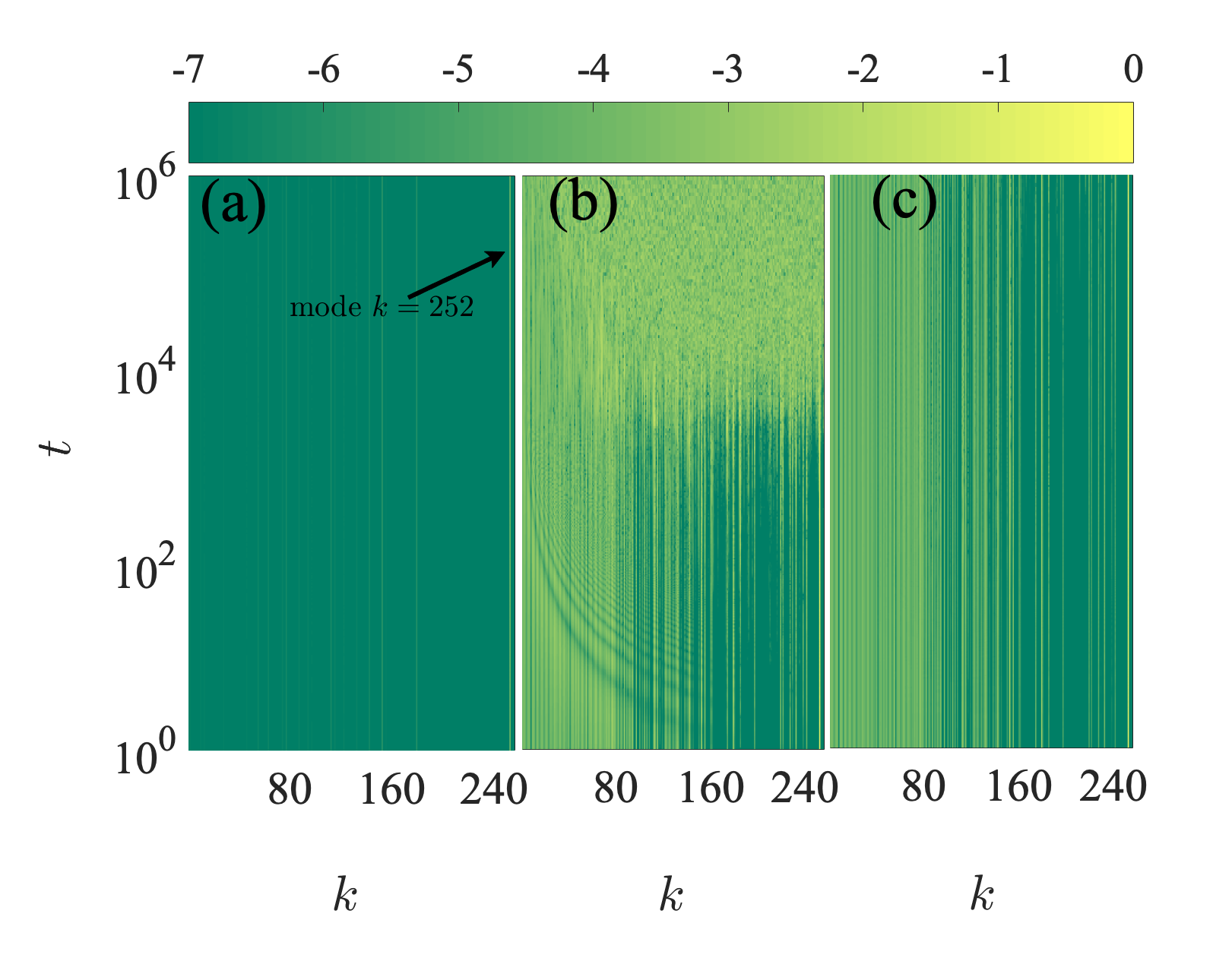}   
\end{center}
\caption{(color online) Spatiotemporal evolution of normal mode energy content after a momentum single site initial excitation for the chain with non-central forces (a) $H_3$ with energy $H_3 = 10^{-4}$, (b) $H_3$ with energy $H_3 = 0.2$ and (c) $H_4$ with energy $H_4 = 0.2$. The color bar is in logarithmic scale. }
\label{fig5}
\end{figure}
To better understand the fate of single site initial excitations, let us discuss the evolution of normal modes. To do so, we transform the linear part of the Hamiltonian (\ref{eqH1}), and cast it in terms of the system's normal modes. At any time ${t}$, we define the velocity vector 
$\vec{V}(t) = [\dot{u}_1(t),\ldots, \dot{u}_{N}(t)]^{T}$. The projection of this vector onto the system's normal modes is given by $ \vec{T} = \mathbf{A}^{-1} \vec{V}(t) $ with matrix $ \mathbf{A}$ having the lattice eigenvectors as columns. In the same manner, the displacement vector $\vec{U}(t) =[u_1(t),\ldots,u_N(t)]^{T}$ is projected onto normal modes to yield $ \vec{R} = \mathbf{A}^{-1} \vec{U}(t) $. $T_k$ and $R_k$ respectively are elements of the projection vector corresponding to normal mode momenta and positions of the $k$th mode ($k=1,...,N$). Thus the total energy of the linear modes is given as 
\begin{equation} \label{nm}
 H =\sum_{k=1}^{N} E_k= \sum_{k=1}^{N} \left( \frac{ T_k^2}{2} + \frac{ \omega^2_k R^2_k }{2} \right).
 \end{equation}
We first consider the spatiotemporal evolution of the linear modes after a single site momentum excitation of the system has been implemented for the various nonlinearities. The  evolution of normal modes for $H_3$ is shown in Fig.~\ref{fig5} for both the near linear regime [panel (a)] as well as the nonlinear regimes [panel (b) and (c)]. We shift between these regimes by changing the system's energy. 

More specifically, in Fig.~\ref{fig5}(a), we depict results for the evolution of $H_3$ in normal mode space after a single site momentum initial excitation for the near linear regime where mainly the localized mode $k = 252$, is significantly excited ($\approx 90\% $ of the total energy), although some other normal modes are weakly excited as well ($\approx 10\% $ of the total energy). However, no coupling of the normal modes takes place during the dynamical evolution. On the other hand, an increase in energy entails that the system becomes more nonlinear. We depict results for this nonlinear regime in Fig.~\ref{fig5}(b). There we see that the nonlinearity due to the cubic potential is able to facilitate coupling of the linear normal modes until the modes are well mixed especially after $t \gtrsim 10^{4}$. We shall investigate further this mixing of linear eigenmodes in Sub-section~\ref{ssec31} where we will discuss the route to energy equipartition. Interestingly enough, the quartic nonlinearity ($H_4$) does not appear to facilitate normal mode mixing as depicted by the results in Fig.~\ref{fig5}(c) which show that, although other normal modes besides mode $k =252$ are initially excited, they do not mix at all. The results for $H_{34}$ (not shown here) are similar to those for $H_3$. 
\subsection{Energy delocalization}
\begin{figure}[t!]
\begin{center}
\includegraphics[width=0.9\textwidth,height=0.5\textheight,keepaspectratio]{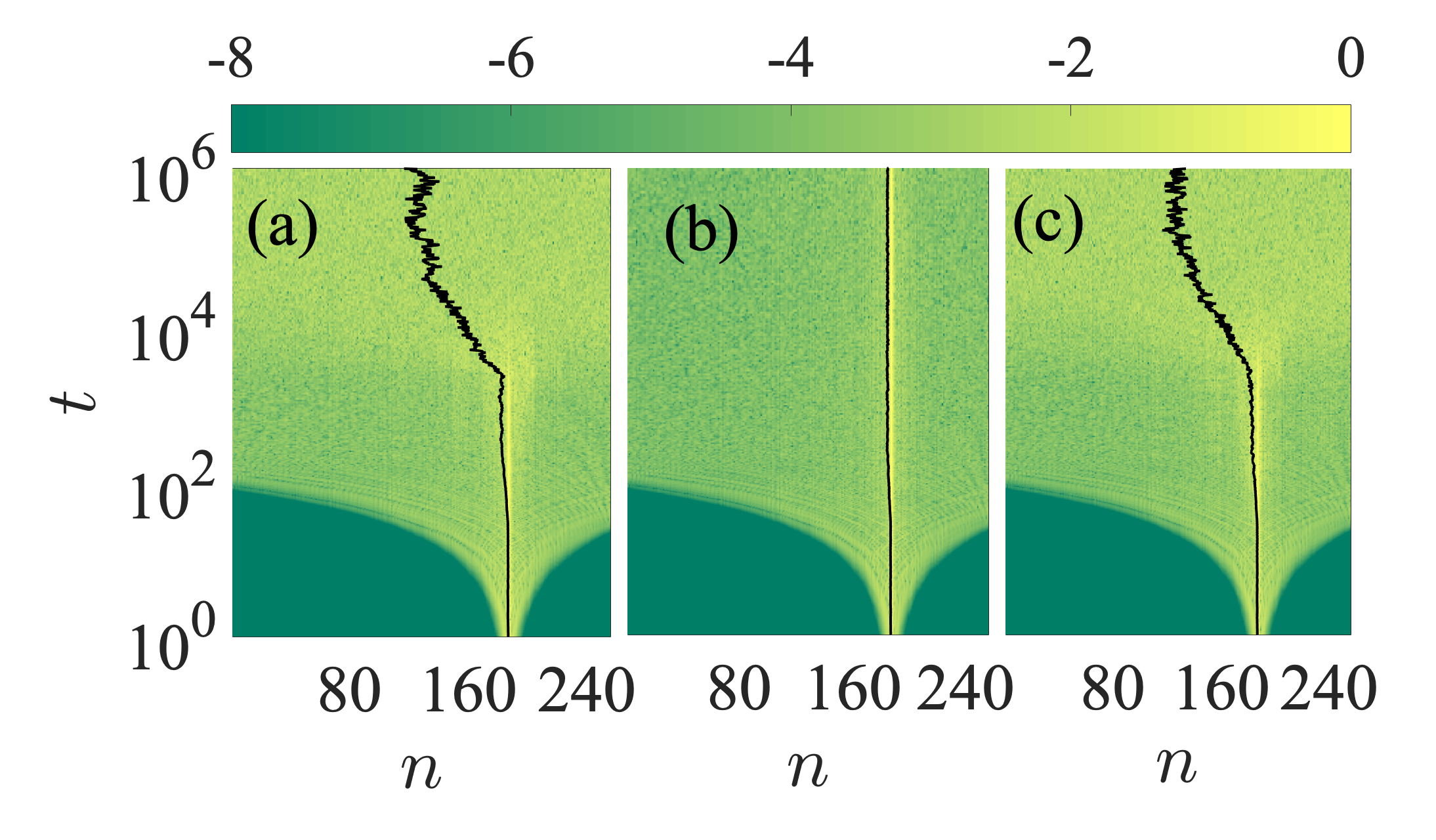}  
\end{center}
\caption{(color online) Spatiotemporal evolution of the energy density after momentum single site initial excitation. Results are shown for systems (a) $H_3$ (b) $H_4$ and (c) $H_{34}$. The total system energy for all three cases is $H = 0.2$. The black curves in all plots depict the mean position of the energy density. The used color bar is in logarithmic scale such that yellower (lighter) regions have more energy than greener (darker) regions.}
\label{fig6}
\end{figure}
We now turn our attention onto another issue and investigate how the initially localized energy spreads in time to the rest of the lattice sites when the system is initialized by a single site momentum excitation. In particular, we follow the lattice dynamics for initially localized energy excitations for systems modeled by the Hamiltonians $H_3$, $H_4$ and $H_{34}$ for the same value of total energy. More specifically, we set $H = 0.2$ and we show results for momentum single site initial excitations in Fig.~\ref{fig6} for the three types of nonlinearities. In Fig.~\ref{fig6}(a), we show the spatiotemporal evolution of the initially localized wave-packet when the system is described by $H_3$. The results in Fig.~\ref{fig6}(a) show the delocalization of the wave-packet which eventually leads to energy spreading to the whole lattice. This happens especially for the time interval $t\gtrsim 10^{4}$. Results for the dynamics when the system is governed by $H_4$ are depicted in Fig.~\ref{fig6}(b) showing that the wave-packet remains localized at least for the duration of our simulation. For a combination of these two types of nonlinearities, i.e., the cubic and quartic nonlinearity, the dynamical behavior again reverts to showing energy spreading for the initially localized wave-packet around site $n= 187$ [Fig.~\ref{fig6}(c)]. To ascertain the energy localization we integrate the systems for the three cases up to $t = 10^6$ time units, which is well beyond the times we initially observe energy spreading. With this in mind we set the final integration time at $t = 10^6$ for all our simulations.

In order to quantify the energy localization, we consider the temporal evolution of the participation number $P$ given by Eq.~(\ref{eqP}) for the three cases. We find that the nonlinear dynamics in cases where the cubic nonlinearity is present (i.e., for the cases of $H_3$ and $H_{34}$), exhibit saturation of the participation number to $P \approx 125$ for $t \gtrsim 10^4$. This is depicted by the blue and green curves in Fig.~\ref{fig7}(a) and the saturation of $P$ is clearly observed by the flattening of these two curves, especially during the last two decades of the evolution. However, for $H_4$, where only the quartic nonlinearity is present, the dynamics reveal localized behavior with no energy spreading as depicted by a practically flat curve of $P$ in Fig.~\ref{fig7}(a) (red curve) [$P(t) \approx 2$]. We believe that this absence of wave-packet spreading is possibly due to the non-excitation of other modes [see Fig.~\ref{fig5}(c)] which does not permit the spreading of energy to the rest of the lattice. It is interesting to note that this behavior is in contradiction to what is observed in the two versions of the homogeneous FPUT model, where it is known that the $\alpha$-FPUT system (cubic nonlinearity) thermalizes much later than the $\beta$-FPUT one (quartic nonlinearity) \cite{57}. Even though our system is disordered, $H_3$ and $H_{34}$ reach equipartition faster than $H_4$.
A combination of the two kinds of potentials ($H_{34}$) shows that the dynamics is strongly dominated by the cubic potential as its behavior is similar to the one observed for the $H_{3}$ model. In fact, there are not many differences between the dynamics of $H_{3}$ and $H_{34}$. In fact by comparison of Figs.~\ref{fig6}(a) and (c) as well as the blue and green curves in Fig.~\ref{fig7}(a) it is clear that there are not many differences between the dynamics of $H_3$ and $H_{34}$.

\subsection{Energy equipartition} \label{ssec31}

\begin{figure}[h!]
\begin{center}
  \includegraphics[width=0.45\textwidth,height=0.45\textheight,keepaspectratio]{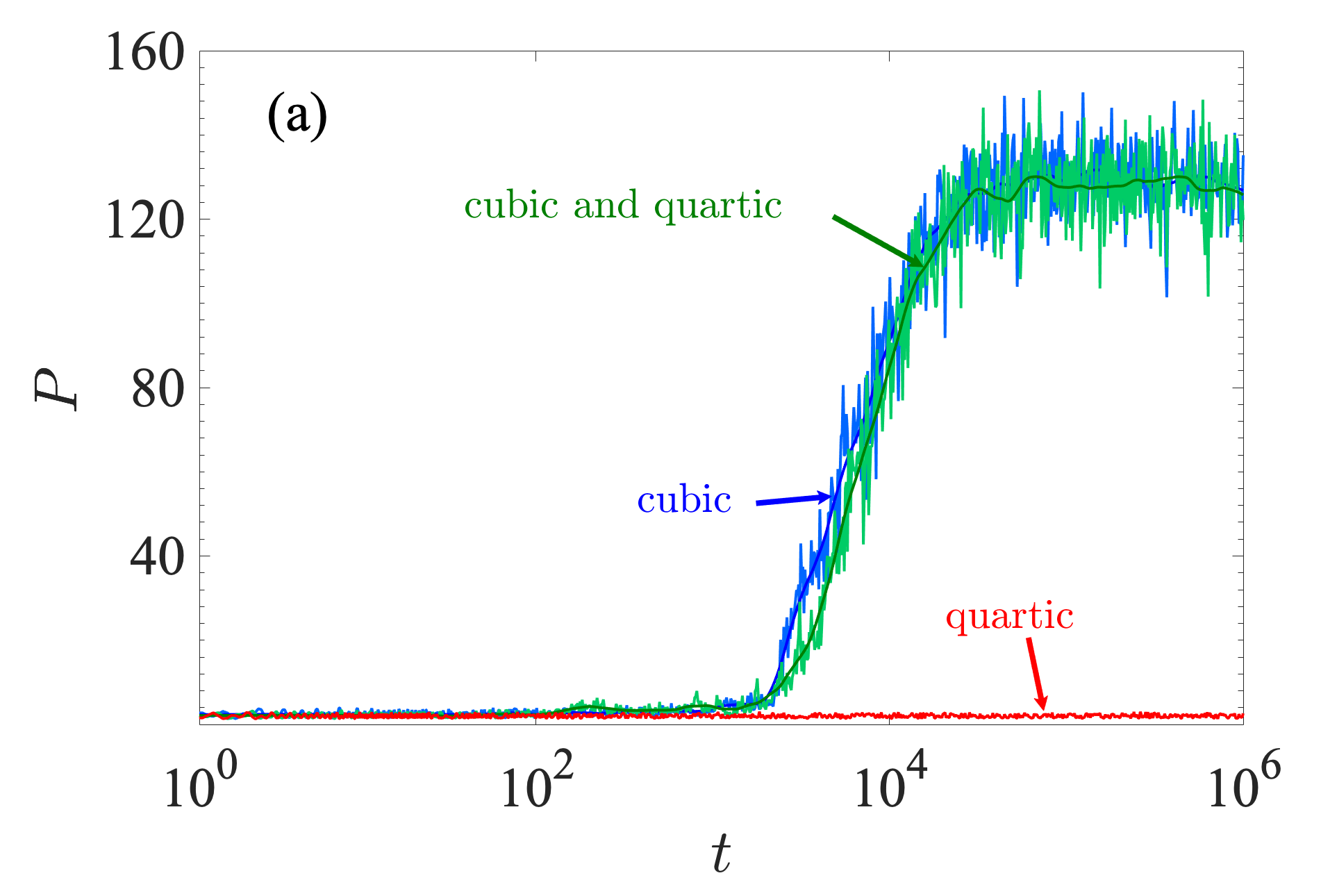} 
  \includegraphics[width=0.45\textwidth,height=0.45\textheight,keepaspectratio]{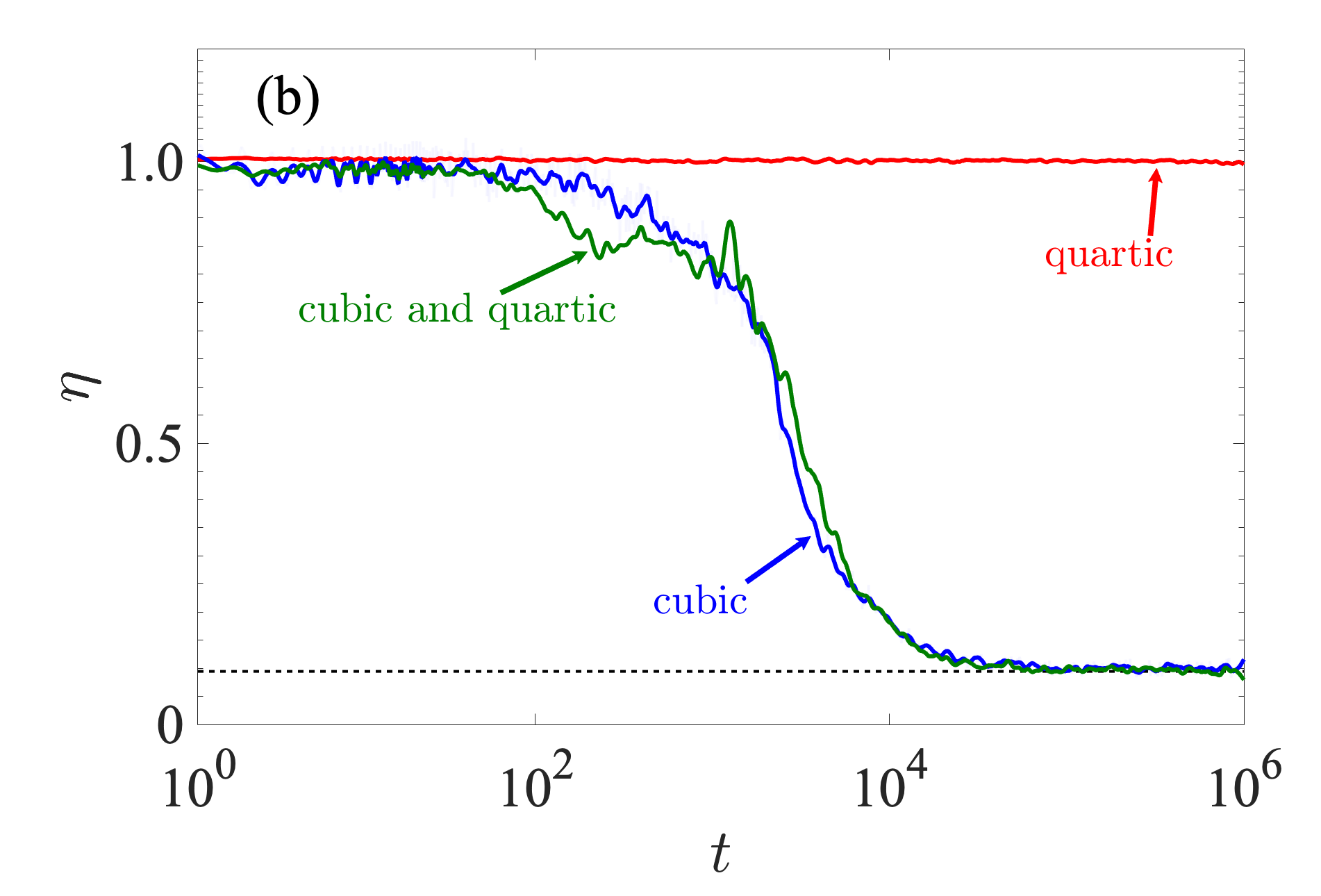}  
\end{center}
 \caption{(color online) Time evolution of (a) $P$~(\ref{eqP}), and (b) $\eta $~(\ref{eqN}); for systems $H_3$ (blue curves), $H_4$ (red curves) and $H_{34}$ (green curves). The total system energy for all three cases is $H = 0.2$. The dashed horizontal line in (b) indicates the analytical value of $\langle \eta \rangle$ when energy equipartition is reached.}
\label{fig7}
\end{figure}
It is also interesting to determine whether the system eventually reaches energy equipartition for the three types of nonlinearities. In order to do so, we numerically compute the so called ``spectral entropy'' by monitoring
the corresponding time evolution of the normal modes \cite{47}. To this end we write the weighted harmonic energy of the $k$th mode as $v_k = E_k/ \sum^{N}_{k=1} E_k$ where $E_k$ (\ref{nm}) is the $k$th linear mode energy. Then the spectral entropy at time $t$ is computed as
\begin{equation} \label{eq31}
S(t) = - \sum^N_{k=1} v_k(t) \ln v_k(t), 
\end{equation}
where $0<S\leq S_{max}=\ln N$.
However, the normalized spectral entropy $ \eta (t) $ in that case is given as
\begin{equation}\label{eqN}
\eta (t) = \frac{ S(t ) - S_{max} }{S(0) - S_{max}}.
\end{equation}
Thus, through such normalization, $0 \leq \eta  \leq 1$. When $\eta $ is close to one, the dynamics
do not substantially deviate from the initially excited mode(s). However,
as more modes are excited, $\eta$ decreases and approaches
zero. For a system at equipartition, a theoretical prediction
for the mean entropy $\eta$ exists, which assumes that the
modes follow a Gibbs distribution when the
nonlinearity is weak. The analytical form of the mean entropy
$\langle \eta \rangle _{an} $ in that case is given as 
\begin{equation} \label{eqentr}
\langle \eta  \rangle_{an}  = \frac{ 1 - C }{\ln N  - S(0)},
\end{equation}
with $C \approx 0.5772$ being the Euler constant \cite{48,49}.
In Fig.~\ref{fig7}(b), we show the time evolution of the spectral entropy $\eta$ in order to determine how the system approaches energy equipartition. When $\eta$ is equal to the analytically predicted value of Eq.~(\ref{eqentr}), we conjecture that energy equipartition is achieved. Again a similar behavior, in terms of how energy equipartition is approached, is observed between $H_3$ and $H_{34}$ where the cubic nonlinearity is active. This is shown in Fig.~\ref{fig7}(b) where there is close resemblance between the results for $H_3$ and $H_{34}$, indicated by the blue and green curves. These curves also show that energy equipartition (indicated by $ \eta \approx \langle \eta  \rangle_{an} = 0.0930 $) is achieved at around the same time as energy delocalization. For the case where the system only has a quartic nonlinearity ($H_4$), we find no energy equipartition, as shown by the corresponding time evolution of $\eta$, which remains fairly constant at almost $\eta = 1$ in Fig.~\ref{fig7}(b).

\section{Chaotic behavior} \label{sec4}
Chaoticity is another aspect that is interesting to study about the dynamics of the system. To study chaos, in this work we utilize a  
global indicator of chaos called the maximum Lyapunov exponent (mLE), which we compute using the so-called standard method \cite{50,51,52}. The finite time mLE (ftmLE) is defined as 
\begin{figure}[ht!]
\begin{center}
  \includegraphics[width=0.65\textwidth,height=0.65\textheight,keepaspectratio]{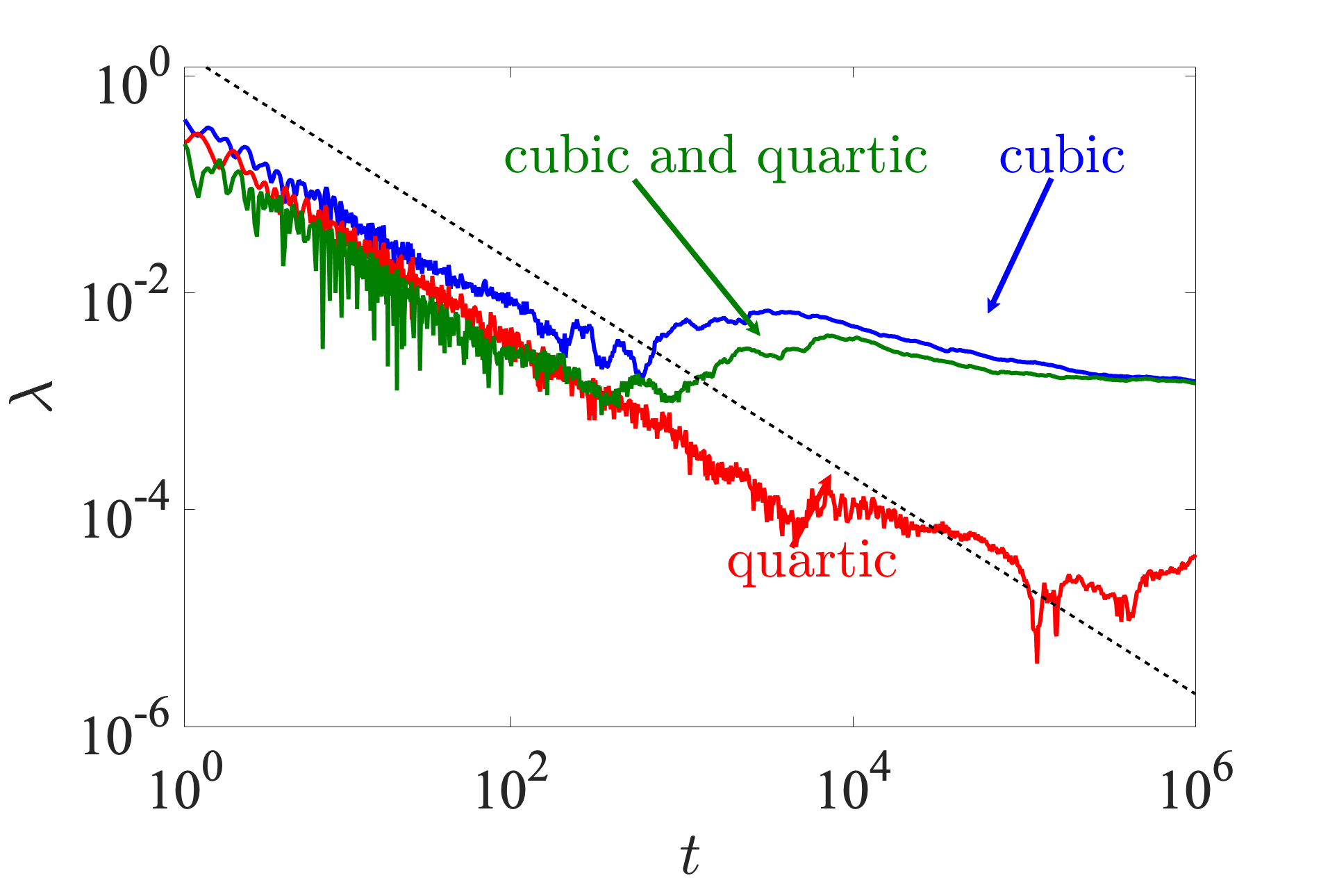} 
\end{center}
\caption{(color online) Time evolution of $\lambda $~(\ref{eq41}) for systems (i) $H_3$ (blue curve) (ii) $H_4$ (red curve) and (iii) $H_{34}$ (green curve) for the same momentum initial excitation discussed in Section~\ref{sec3}. The dashed diagonal line guides the eye for slope $-1$.}
\label{fig8}
\end{figure}
\begin{equation} \label{eq41}
\lambda(t) = \frac{1}{t} \ln \frac { || \mathbf{w} (t) || }{ || \mathbf{w} (0) ||},
\end{equation}
where $\mathbf{w} (t) $ is a vector of small perturbations from the phase space trajectory at time $t$, also called deviation vector, which we denote as 
\begin{equation}
 \mathbf{w} (t)  = [ \delta  u_1 (t), \ldots , \delta  u_N (t), \delta  p_1 (t),  \ldots , \delta  p_N (t)],
\end{equation}
with $\delta  u_n (t)$ and $\delta  p_n (t)$ respectively indicating small perturbations in position and momentum for the $n$th lattice site. The mLE is defined as $ \Lambda = \lim_{t \to \infty} \lambda (t) $. In Eq.~(\ref{eq41}), $ || \cdot ||$ denotes the usual Euclidean vector norm. For chaotic trajectories, $\lambda $ attains a finite positive value, otherwise, $\lambda \propto t^{-1}$ for regular orbits. An efficient and accurate method to follow the evolution of 
 $\mathbf{w} (t)$ is to numerically integrate the so-called variational equations \cite{53}, which govern the vector's dynamics, together with the Hamilton equations of motion using the tangent map method outlined in Refs.~\cite{54,55,56}. 
Let us now investigate the chaotic behavior of the system for the three types of nonlinearity under study. The time evolution of $\lambda$ reveals in all cases a chaotic response, as shown by the ftmLE which deviates from $\lambda (t) \propto t^{-1}$ (see Fig.~\ref{fig8}). In fact, for $H_3$ and $H_{34}$ we attain the same final values of $\lambda$ towards the end of the simulations as shown by the blue and green curves in Fig.~\ref{fig8}. In our study, we also utilize the so-called the deviation vector distribution (DVD) to analyze chaos further. The DVD is able to identify spots in real or phase space which are more sensitive to chaos \cite{29,31,32}. The DVD is written as
\begin{equation} \label{eq42}
\xi_n (t) = \frac{\delta u_n^2 + \delta p_n^2  }{\sum_{n=1}^N (\delta u_n^2 + \delta p_n^2 )}.
\end{equation}
\begin{figure}[b!]
\begin{center}
  \includegraphics[width=0.45\textwidth,height=0.45\textheight,keepaspectratio]{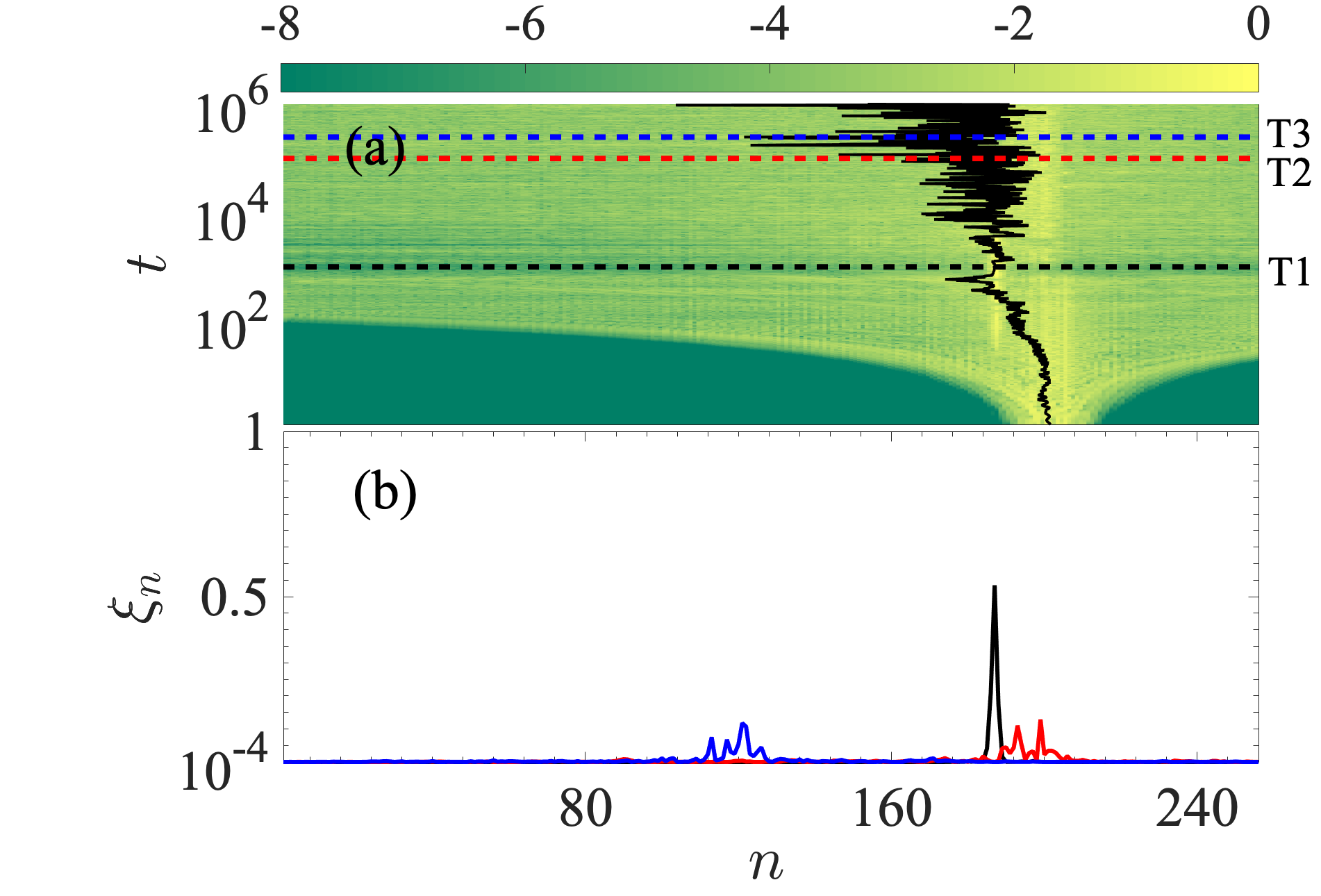} 
 \includegraphics[width=0.45\textwidth,height=0.45\textheight,keepaspectratio]{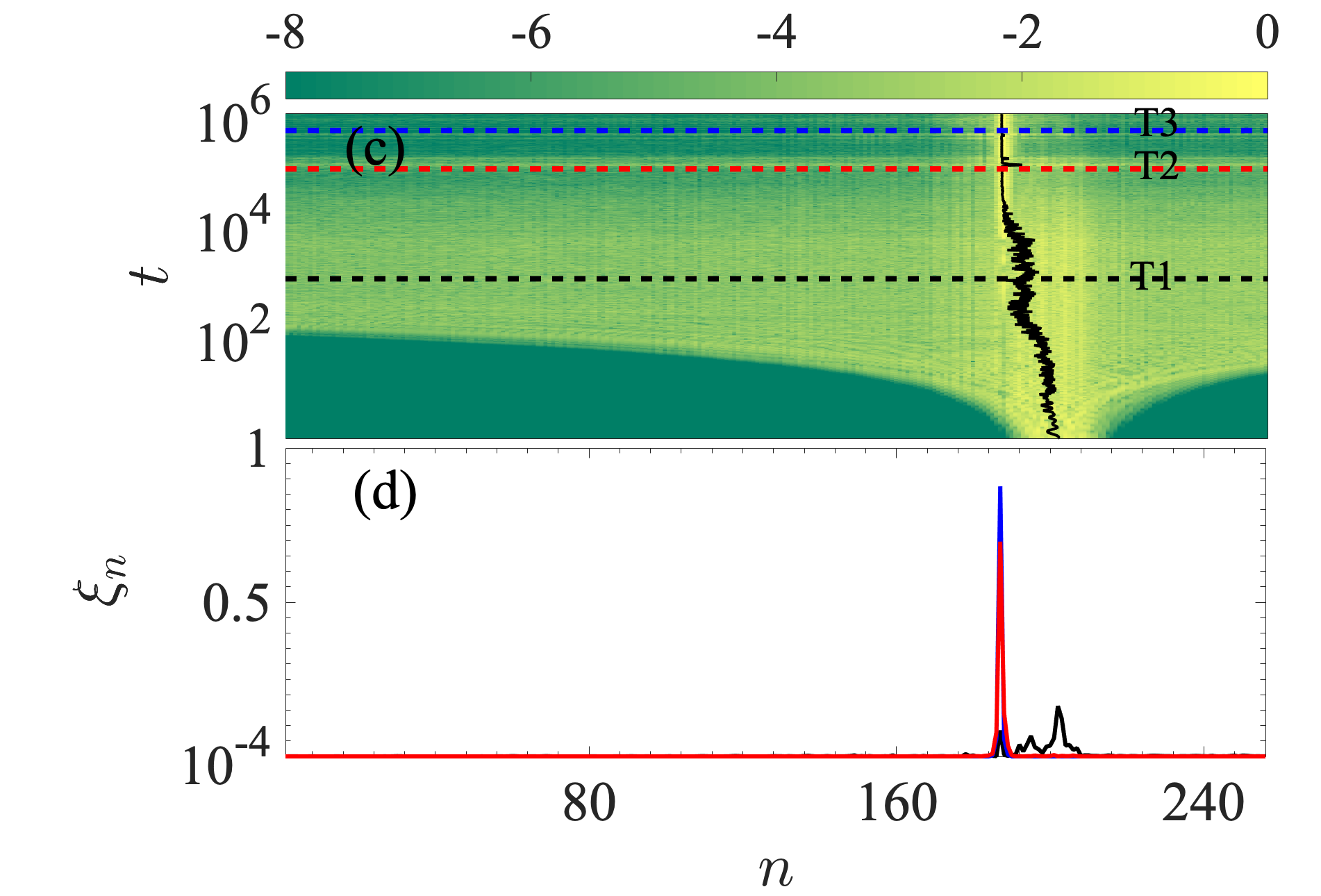} 
\end{center}
 \caption{(color online) (a) Spatiotemporal evolution of the DVD~(\ref{eq42}) and (b) instantaneous snapshots of the normalized DVDs at $t = T1 \approx 2.47 \times 10^3$ (black curve), $t = T2 \approx 2.08 \times10^5$ (red curve) and $t = T3 \approx 2.477 \times 10^5$ (blue curve) respectively corresponding to black, red and blue horizontal dashed lines in (a). The results are for the momentum initial excitation discussed in Section~\ref{sec3} for $H_3 = 0.2$. (c)-(d) are similar to (a)-(b) but for $H_4 = 0.2$. The snapshots were taken at $t =T1  \approx 10^3$, $t = T2 \approx 10^5$ and $t = T3 \approx 5 \times 10^5$. The black curves in (a) and (c) depict the mean position of the DVD. The color bar is in logarithmic scale such that yellower (lighter) regions have more energy than greener (darker) regions.}
\label{fig9}
\end{figure}

We are thus able to describe the spatial distribution of chaos in a dynamical system by making use of the DVD. We now identify the most active chaotic spots in the lattice by following the time evolution of the DVDs. We use as a representative case the results of $H_3$ indicated by the blue curve in Fig.~\ref{fig8}. The spatiotemporal evolution of the DVD is plotted in Fig.~\ref{fig9}(a) and shows that, on average, the mean position of the DVD remains close to the excitation region during the early stages of the dynamics $ t\lesssim 10^3$. Beyond $t \gtrsim 10^3$, the mean position of the DVD  starts to move throughout the chain. This is shown by the erratic motion of the mean position (solid black curve), which implies that there is a chaotic spot randomly moving throughout the lattice. In order to clearly show this behavior, we take some snapshots of the DVD especially towards the end of the simulation ($t \gtrsim 10^5$). These snapshots show peaks at different positions implying that indeed the chaotic spots are moving throughout the chain as shown in Fig.~\ref{fig9}(b). In the case of the combination of cubic and quartic potentials ($H_{34}$) the results are qualitatively similar to those presented in Figs.~\ref{fig9}(a)-(b) hence we do not show them here. The case which has only a quartic potential ($H_4$) is also found to be chaotic, as shown by its $\lambda (t) $ (red curve in Fig.~\ref{fig8}). Upon closer analysis of the DVDs, the chaotic behavior exhibited by $H_4$ is revealed to be different from the behavior already described for the other two types of nonlinearities which feature cubic potentials ($H_3$ and $H_{34}$). In particular, the DVD indicates that chaos remains localized around the excitation region for $H_4$ as is shown by the DVD and the corresponding snapshots in Figs.~\ref{fig9}(c) and (d). Note that $\mathbf{w} (0)$ is centered slightly to the right of the excitation site in Fig.~\ref{fig9}.

\section{Statistical analysis} \label{sec5}
\begin{figure}[t!]
\begin{center}
  \includegraphics[width=0.95\textwidth,height=0.95\textheight,keepaspectratio]{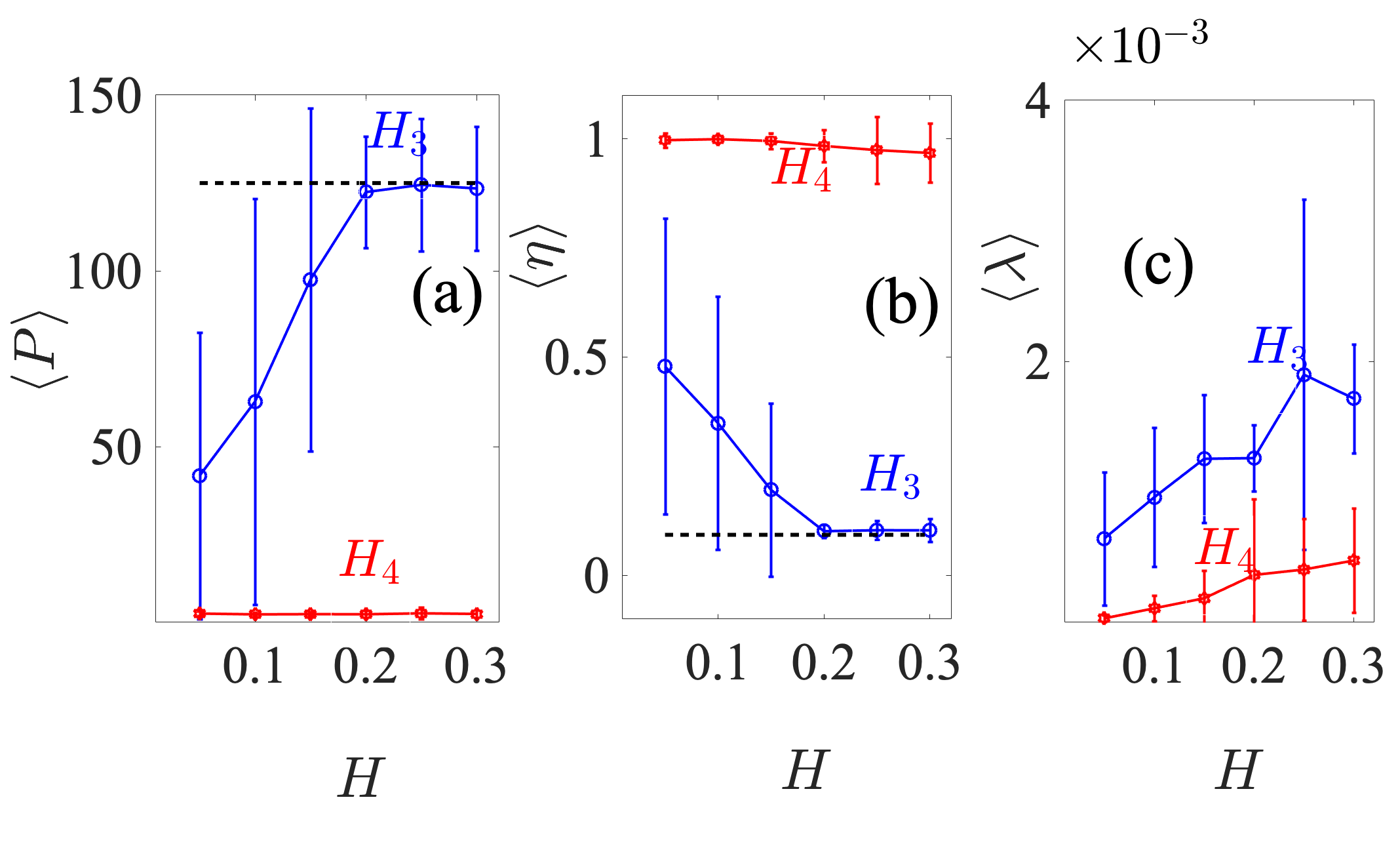} 
\end{center}
\caption{(color online) Mean (a) participation number $ \langle P \rangle $ (b) entropy $ \langle \eta \rangle $ and (c) ftmLE, $ \langle \lambda  \rangle$ for $H_3$ (blue curves) and $H_4$ (red curves) against system's energy $H$, over $30$ disorder realizations for which highly localized modes are taken as initial excitations. The error bars represent one standard deviation.
The dashed line in (a) guides the eye to the level $\bar{P} = 125$, whilst a similar line in (b) guides the eye to the analytical mean entropy at equipartition [Eq.~(\ref{eqentr})].}
\label{fig10}
\end{figure}

We further study how the system behaves statistically by considering different mass disorder realizations for several energy levels from $H = 0.05$ up to $H = 0.3$. In particular, we present results obtained over $30$ realizations at each energy level. The strength of the disorder is such that there is a significant number of strongly localized modes and many of those are in the center (they are located all along the chain) similar to the one already presented in Section~\ref{sec2}. The results for $H_3$ indicate that the mean participation number $ \langle P \rangle $ saturates to $\langle P \rangle \approx 125$ (or $\approx N/2$ ) as the energy is increased for $H_{3}$ [Fig.~\ref{fig10}(a)]. We also find a threshold energy ($H \gtrsim 0.2$) above which, on average, any realization for which a highly localized mode if taken as initial condition leads to this  maximum value of $P \approx 125$. For the quartic potential ($H_4$), we find no energy delocalization and $ \langle P \rangle $ remains very low as the energy is increased as shown by the red curve in Fig.~\ref{fig10}(a) where $ \langle P \rangle \approx 2 $. We studied the system for even higher energy excitations up to $H = 0.9$ and we still found the system's density distribution to remain localized.
The statistical results of the mean spectral entropy, $\langle \eta \rangle$ show contrasting behaviors for $H_{3}$ and $H_{4}$. $H_3$ is found to reach energy equipartition as the predicted analytical value is reached ($\langle \eta \rangle \approx \langle \eta \rangle_{an}$) for what appears to be an energy threshold of around $H \gtrsim 0.2$, as shown by the blue curve in Fig.~\ref{fig10}(b). On the other hand, $H_{4}$ does not reach energy equipartition even at higher energy excitations than those presented Fig.~\ref{fig10}. The results for $H_{4}$ are shown by the red curve in  Fig.~\ref{fig10}(b) for which $\langle \eta \rangle \approx 1$ for all initial energies  indicating no energy equipartition. In fact the smaller error bars on the blue curves for energies $H \gtrsim 0.2$ in both Figs.~\ref{fig10}(a) and (b) validates these conclusions. This result is in some sense the inverse of what was observed for the homogeneous $\alpha$-FPUT model, which attains equipartition later than the $\beta$-FPUT model \cite{57}. In the latter model, one discovered route to equipartition is through modulational instability which has an energy threshold \cite{57,58}. 

As for the system's chaoticity, the final value of the mean ftmLE $ \langle \lambda \rangle $ does not practically depend on the system's total energy; for as depicted in Fig.~\ref{fig10}(c) where $\langle \lambda \rangle$ obtains values around $1.5 \times10^{-3}$ for both $H_{3}$ and $H_{4}$, at least for the energy range considered in this study, although a slight increase of $ \langle \lambda \rangle $ is seen as energy grows. $H_4$ is found to be in general less chaotic than $H_3$ as shown by the lower values of $ \langle \lambda \rangle $ [red curves in Fig.~\ref{fig10}(c)] for $H_4$. Note that the statistical results for $H_{34}$ are qualitatively similar to those for $H_{3}$ presented in Fig.~\ref{fig10}.

 \section{Conclusions} \label{sec6}
In this study, we numerically investigated the dynamics of a disordered $1$D nonlinear lattice model  which is relevant to the study of bending waves, focusing on the energy delocalization and equipartition, as well as the system's chaotic behavior. For this study, we chose a localized wave-packet which is almost similar to a highly localized mode as the initial energy excitation. We considered three kinds of nonlinear configurations of the Hamiltonian: (a) with only a cubic nonlinearity named as  $H_3$, (b) with only a quartic nonlinearity called $H_4$, and finally (c) a combination of the cubic and quartic terms namely $H_{34}$.

The dynamics shows that the energy delocalization strongly depends on the system's type of nonlinearity. A threshold exists beyond which any initial excitation will lead to energy delocalization for the cases of $H_3$ and $H_{34}$ in which the cubic potential dominates the dynamics. The dynamics for $H_{4}$ remains localized throughout the whole simulation even for the large values of the energy. % values we considered here. In fact, even substantially higher energies (up to $H= 0.9$) than those we present in this paper, do not lead to energy delocalization, at least up to the timescales we adopt (i.e., $t \approx 10^6$). 
Similarly, energy equipartition is observed only when the cubic terms are present, even from a statistical viewpoint when the system is studied at different energy levels. 

Additionally, for the same initial conditions, we studied the system's chaotic response using the finite time mLE and the related DVDs. Here all three types of nonlinearities induce chaotic responses. Statistically, the chaotic behavior appears to be slightly stronger when cubic terms are included in the model (systems $H_3$ and $H_{34}$), with  the strength of the chaoticity moderately increasing  as the system energy grows. The chaotic behavior of the systems as analyzed using the DVDs shows both localized and delocalized chaos. Localized chaos is exclusive to $H_4$ even for energy excitations as high as $H = 0.9$. 

We find that our numerical results have not only revealed interesting nonlinear dynamical behaviors but also open the road for further theoretical studies of chains with non-central forces. This is due to the fact that the proposed model suggests fundamentally different dynamics than the ones of well-studied discrete systems like the  DKG and FPUT models, for which we have a better understanding. In particular, our results regarding the very different behavior between the cubic and quartic nonlinearities, poses the question of what are the mechanisms responsible for these observations especially in view of the comparison with the $\alpha$ and $\beta$-FPUT models. Along the same lines, an interesting question is how  the energy spreading is obtained in infinite chains and whether the increase of nonlinearity leads (or not) to  subdiffusive behaviors.

Our work furthermore contributes to the debate about the potential localization or spreading of initially localized excitations in disordered nonlinear systems when such processes are juxtaposed to the system's chaotic strengths. 

\section*{Acknowledgements}
Ch.~S.~thanks the Universit\'e du Mans for its hospitality during his visits when part of this work was carried out. We also thank the Centre for High Performance Computing (\url{https://www.chpc.ac.za}) for providing computational resources for performing significant parts  of this paper's computations. A.N. acknowledges funding from the University of Cape Town (University Research Council, URC) postdoctoral Fellowship grant and the Oppenheimer Memorial Trust (OMT) postdoctoral Fellowship grant

\end{document}